\documentclass{mem}
\usepackage{natbib}

\usepackage{balance}
\usepackage{graphicx}
\usepackage{flushend}
\usepackage{amsmath}
\usepackage{txfonts}
\usepackage{bbold}
\usepackage[breaklinks,pdftex]{hyperref}
\idline{75}{282}

\begin{document}

\title{
An orphan flare from a plasma blob crossing
the broad-line region ?}

\author{
S. \,Le Bihan\inst{1}  \and A. \, Dmytriiev\inst{2} 
\and A. \, Zech\inst{1} 
          }

\institute{
LUX, Observatoire de Paris, Université PSL, Sorbonne Université, CNRS, 92190 Meudon, France \email{sebastien.le-bihan@obspm.fr}
\and
Centre for Space Research, North-West University, Potchefstroom, 2520, South Africa\\
}
\authorrunning{Le Bihan}

\titlerunning{Orphan flare from a blob crossing the BLR}

\date{Received: XX-XX-XXXX (Day-Month-Year); Accepted: XX-XX-XXXX (Day-Month-Year)}

\abstract{
The blazar 3C\,279 is well known for its prolific emission of rapid flares. A particular event occurred on 12/20/2013, exhibiting a large flux increase with a doubling time scale of a few hours, a very hard $\gamma$-ray spectrum, and a time-asymmetric light curve with slow decay, but no significant variations detected in the optical range.
We propose a novel scenario to interpret this flare, based on two emission zones, a stationary blob and a moving plasma blob.
The stationary blob, located within the BLR, accounts for the low-state emission. The moving blob decouples from the stationary zone, accelerates and crosses the BLR. The high-energy flare is attributed to the variable external Compton emission as the blob moves through the BLR, while variations in the synchrotron emission are negligible. Our interpretation differs from previous interpretations by attributing the flare to the bulk motion and geometry of the external photon fields, without invoking varying electron injection.
%{\bf NOT MORE THAN 1000 characters including spaces are allowed, please double-check before submit the paper.}
}
\maketitle{}

\section{Introduction}

Flat Spectrum Radio Quasars (FSRQs) are high-luminosity Active Galactic Nuclei (AGNs) with a relativistic jet pointed toward the Earth. Jets are thought to be responsible for the broad-band emission of the FSRQ, with the synchrotron emission mechanism that emits from radio waves up to X-rays, and the inverse Compton (IC) mechanism that emits from X-rays up to very-high-energy $\gamma$-rays. The IC can be caused by the upscattering of photons coming from the synchrotron emission (synchrotron-self-Compton - SSC), or by the upscattering of photons from an external radiation field, the source of which can be the accretion disk, the broad line region (BLR) or the dusty torus of the AGN (external Inverse Compton - EIC). The latter is an important mechanism for the FSRQs because of their high-luminosity accretion disk and thus strong external photon fields. The synchrotron emission and the IC produce very distinguishable bumps in the spectral energy distribution (SED), while the EIC and SSC can or cannot be distinguished depending on the source.

On December 20$^{th}$ 2013 (MJD 56646) the FSRQ 3C\,279 ($z = 0.536$, well known for its prolific emission of rapid flares) was subject to a particular type of flare : a short ($\approx 12 \ hours$) and strong flare was recorded \citep{Hayashida_2015} in the high-energy $\gamma$-ray band ($20 \ MeV - 300 \ GeV$), but not in the optical and infrared bands. During the flare period, the source was not observed in the X-ray band, leaving us without information in this energy range.

We describe this apparent orphan flare using the time-dependant EMBLEM code \citep{dmytriiev2021}, with a scenario that differs from previous scenarii (\citealt{Hayashida_2015,Lewis2019,Paliya2016}) in that it does not involve any acceleration or additional injection of the plasma electrons within the jet, but it is a consequence of the bulk
motion and geometry of the external photon fields

\section{Multiwavelength observation of 3C\,279 orphan flare}

Data available over three days before the flare show 3C 279 in a quiescent state. We refer to this state as Period LS (MJD 56643-56645). The flare period is referred to as Period F (MJD 56646). 

During Period LS, 3C\,279 was observed in $\gamma$-rays by {\it Fermi}-LAT, in X-rays by {\it NuSTAR} and {\it Swift}-XRT, in UV and optical by {\it Swift}-UVOT, in optical and near infrared by the Kanata Telescope and the Small and Medium Aperture Research Telescopes (SMARTS) and in radio by the Submillimeter Array (SMA).
Fewer telescopes observed 3C\,279 during Period F because of the short duration of the flare. Thus for this period we only have data from Fermi-LAT, the Kanata telescope and SMARTS. 

\begin{figure}
\resizebox{\hsize}{!}{\includegraphics[clip=true]{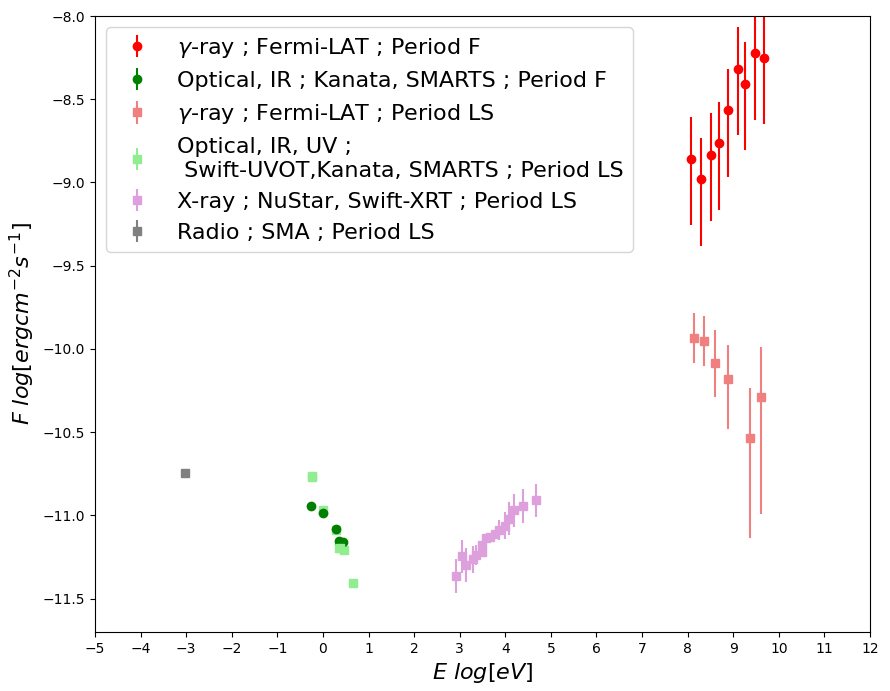}}
\caption{
\footnotesize
Broadband spectral energy distribution of 3C\,279 for the Period LS (low-state of 3C\,279) and the Period F (flaring state of 3C\,279). Error bars are 1$\sigma$ statistical errors. Data points are from \citet{Hayashida_2015}.}
\label{fig:seddata}
\end{figure}

\begin{figure}
\resizebox{\hsize}{!}{\includegraphics[clip=true]{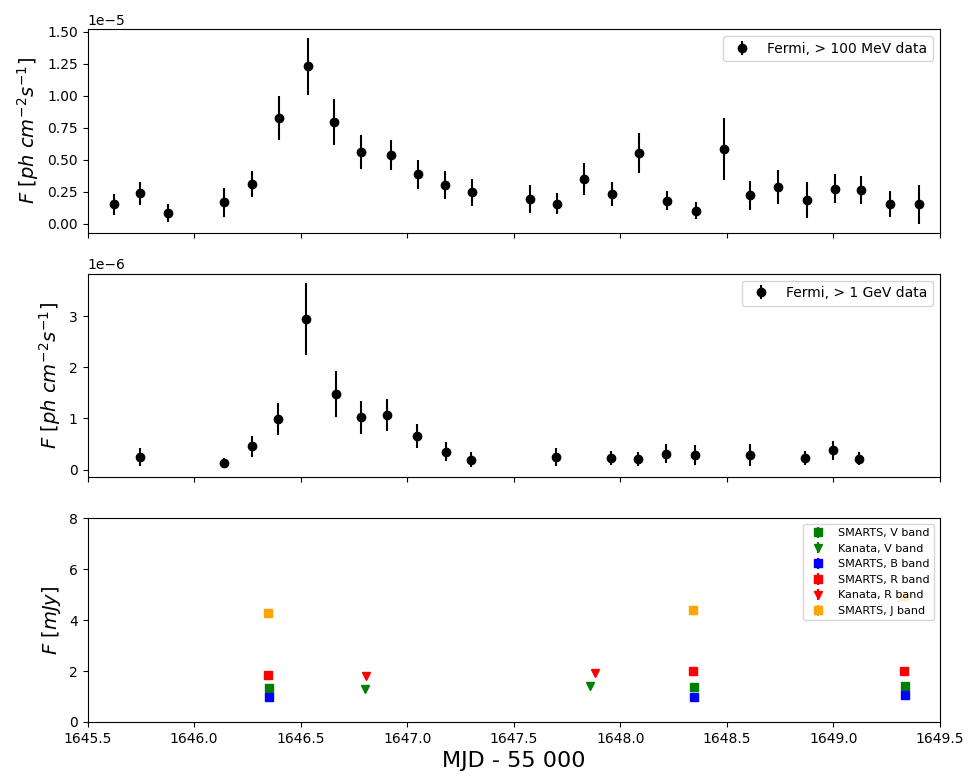}}
\caption{
\footnotesize
$\gamma$-ray, optical, and infrared light curves of 3C 279. The top and middle panel shows the integrated photon flux for energies $> 100$ MeV and $> 1$ GeV (192 minutes bin). The bottom panel displays the effective flux density in optical and infrared bands. Error bars are $1\sigma$ statistical errors. Data points are from \citet{Hayashida_2015}.}
\label{fig:lcdata}
\end{figure}

The SED and light curve data are taken from \citet{Hayashida_2015} and are plotted in Figures \ref{fig:seddata} and \ref{fig:lcdata}. These data exhibit a significant flux variation in the $\gamma$-ray band but not in the optical and infrared bands. The flux variation is described by a large flux increase with a doubling time scale of a few hours, a very hard $\gamma$-ray spectrum and a time asymmetry with a slow decay. Only two flares with these characteristics had been observed so far: the present one and that of PKS\,1510-089 on MJD 55854 \citep{Nalewajko2013}.

\section{Description of our scenario}

A relativistic jet is launched from the black hole at an angle $\theta_j$ with the line of sight. The plasma is accelerated from the base of the jet up to a certain distance from the black hole. Within the jet, two compact zones ("blob 1" and "blob 2") are responsible for the emissions of the FSRQ. Blob 1 is a stationary emission region, traversed by
a bulk plasma flow that has a Lorentz factor $\sim 11$. Blob 2 is accelerated as it moves within the jet and its Lorentz factor is given by \citep{ghisellini2009}:
\begin{equation}
	\Gamma_{blob} = \min\left(\Gamma_{max}, \left(\frac{R_{b-BH}}{3R_S}\right)^{1/2}\right)
	\label{acc_blob}
\end{equation}
with $R_{b-BH}$ the distance between the blob and the black hole, $R_S$ the Schwarzschild radius of the black hole and $\Gamma_{max}$ corresponding to the maximal velocity of the jet. The acceleration of the plasma flow is ascribed to differential collimation of a magnetised jet.

Accretion disk, BLR and dusty torus each provide an external photon field. The field from the disk is described as in \citet{ghisellini2009} with a multi-temperature black body. The BLR is emitting a spectrum of lines that are a result of the re-processing of a fraction $f_{BLR}$ of the disk radiation. It is described as a spherical shell with decreasing matter density at radii above the inner radius $R_{BLR}$. In the blob frame, its radiation energy density is \citep{hayashida2012} :
\begin{equation}
	U'_{BLR}(\epsilon',R_{b-BH}) = \frac{L'_{BLR}(\epsilon')\Gamma_{blob}^2}{3\pi R_{BLR}^2 c \bigg[1+\bigg(\frac{R_{b-BH}}{R_{BLR}}\bigg)^{\beta_{BLR}}\bigg]}
	\label{rad_BLR}
\end{equation}
where the prime designates the blob frame, $L_{BLR} = f_{BLR} L_d$ is the fraction of the disk luminosity re-emitted by the BLR, $R_{BLR}$ ($= 0.1 \sqrt{L_{d,46}} \ pc = 3.1 \times 10^{17} $\,cm in our model) is the inner radius of the BLR shell, $\Gamma_{blob}$ is the Lorentz factor of the blob and $\beta_{BLR}$ is the power law index of the BLR density profil. Emission from the dusty torus is included in the code, but is negligible for the present model.

In our two-zone model, blob 1 is stationary within the BLR and accounts for the AGN's quiescent state (Period LS). Blob 2, responsible for the flare emission (Period F), is accelerated from the jet base up to the BLR shell. Inside the shell, as blob 2 accelerates, the BLR radiation energy density in the blob frame increases with the distance from the black hole as (eq. \eqref{acc_blob} and eq. \eqref{rad_BLR}):
\begin{equation}
    U'_{BLR} \propto R_{b-BH}
    \label{eq:inside_BL}
\end{equation}
The maximum bulk Lorentz factor is set to $\Gamma_{max} = 30$, such that acceleration stops around $R_{b-BH} = 4.0 \times 10^{17}$\,cm. After that, as blob 2 moves at a constant velocity, the energy density evolves as:
\begin{equation}
    U'_{BLR} \propto R_{b-BH}^{-\beta_{BLR}}
    \label{eq:outside_BL}
\end{equation}
As blob 2 moves along the jet, the initial increase in the BLR radiation energy density is followed by a decrease after acceleration stops, leading to a corresponding rise and fall in the EIC emission and thus to a high-energy flare.

The emission is considered to be entirely due to the population of relativistic electrons or pairs in the two plasma blobs. The evolution of the electron distribution within the blob is given by the following Fokker-Planck equation (\citealp{kardashev1962,tramacere2011}):
\begin{equation}
\begin{split}
    \frac{\partial N_e(\gamma,t)}{\partial t} = 
    \frac{\partial}{\partial \gamma} \bigg[\bigg(\frac{1}{t_{cool}(\gamma,t)}+\frac{1}{t_{ad}}\bigg) \gamma N_e(\gamma,t)\bigg] \\
    - N_e(\gamma,t)\bigg(\frac{1}{t_{esc}(t)}+\frac{3}{t_{ad}}\bigg)
    + Q_{inj}(\gamma,t)
\end{split}
\end{equation}
with $\gamma$ the Lorentz factor of the electrons, $t_{cool}$ the characteristic cooling timescale due to synchrotron and IC emission \citep{dmytriiev2021}, $t_{ad}$ the adiabatic cooling timescale due to the expansion of the blob \citep{tramacere2022}, and $t_{esc}$ the escape timescale of the electrons leaving the blob. The injection rate $Q_{inj}$ is given by:
\begin{equation}
	Q_{inj}(\gamma,t) = N \ \ \bigg(\frac{\gamma}{\gamma_{pivot}}\bigg)^{\alpha_{inj}}
	\label{inj_eq}
\end{equation}
for $\gamma_{min} \leq \gamma \leq \gamma_{max}$, and accounts for the continuous injection of electrons in the blob with the shape of a power law spectrum of norm $N$ and index $\alpha_{inj} < 0$. 

We assume a constant injection rate and an opening angle of the jet of $0.5^{\circ}$ as observed by \citet{opening_angle}. As blob 2 is subject to adiabatic expansion, its radius $R_b$ and its tangled magnetic field $B$ are evolving with time \citep{tramacere2022}. For blob 1, these parameters are constant and its electron population is in a steady state.

All values of the different parameters used are listed in Tables \ref{tab:tab1} and \ref{tab:tab2}. 

\begin{table}[ht]
\caption{Parameters used in the simulation of emission from the two blobs.}
\label{tab:tab1}
\centering
\begin{tabular}{lcc}
    \hline
    \textbf{Parameter} & \textbf{Blob 1} & \textbf{Blob 2} \\
    \hline
    \multicolumn{3}{c}{\textbf{Blob characteristics}} \\
    \hline
    Type & Stationary & Accelerating \\
    $B$ ($G$) & $3.2$ & $[0.01, 0.0086]$ \\
    $R_{b-BH}$ ($cm$) & $0.5 \times 10^{17}$ & $[0.5, 50] \times 10^{17}$ \\
    $R_b$ ($cm$) & $3.2 \times 10^{15}$ & $[2.4, 2.8] \times 10^{16}$ \\
    \hline
    \multicolumn{3}{c}{\textbf{Injection spectrum}} \\
    \hline
    Type & PL & PL \\
    $N$ ($cm^{-3}s^{-1}$) & $1.0 \times 10^{-2}$ & $2.5 \times 10^{-4}$ \\
    $\gamma_{\mathrm{pivot}}$ & $215$ & $200$ \\
    $\alpha_{\mathrm{inj}}$ & $-3.5$ & $-3.5$ \\
    $\gamma_{\mathrm{min}}$ & $500$ & $1300$ \\
    $\gamma_{\mathrm{max}}$ & $10^5$ & $10^5$ \\
    \hline
\end{tabular}
\end{table}

\section{Results and discussion}

\begin{figure}
\resizebox{\hsize}{!}{\includegraphics[clip=true]{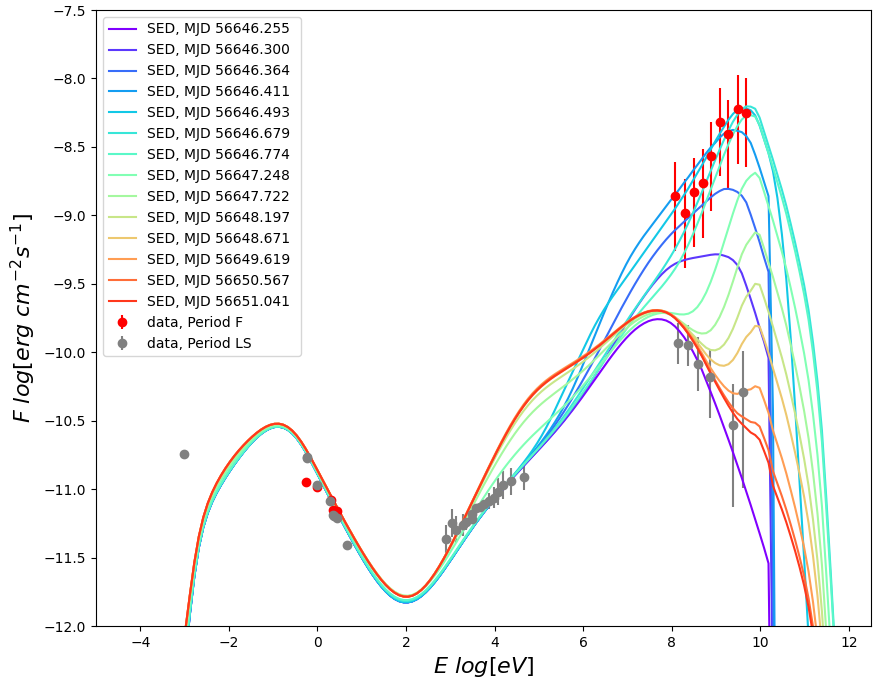}}
\caption{\footnotesize
Simulated SED of both blob 1 and blob 2 emissions. Blob 1 emission explains the data of Period LS. Blob 2 IC-EC emission explains the high-energy data ($E \gtrsim 0.1 \ GeV$) of Period F.}
\label{fig:sedmodel}
\end{figure}

The time evolution of the particle populations as well as the resulting SED and light curve due to the synchrotron and IC emission is computed with the EMBLEM code. Figure \ref{fig:sedmodel} shows the combined SED of both emission regions. While blob 1 dominates the synchrotron emission at all times, due to a substantially higher magnetic field, and reproduces the high-energy bump during the low state, emission from blob 2 leads to a flare at high energies.
\begin{table}[ht]
\caption{AGN parameters used for the model.}
\label{tab:tab2}
\centering
\begin{tabular}{lc}
\hline
\textbf{Parameters} & \textbf{Values} \\
    \hline
    Redshift ($z$) & $0.536$ \\
    Initial Doppler factor ($\delta_i$) & $18.8$ \\
    Final Doppler factor ($\delta_f$) & $30$ \\
    Jet angle ($\theta_j$, degrees) & $1.91^\circ$ \\
    Black hole mass ($M_{BH}$, $M_\odot$) & $0.5 \times 10^9$ \\
    Disk luminosity ($L_d$, erg/s) & $1.0 \times 10^{46}$ \\
    BLR fraction ($f_{BLR}$) & $0.1$ \\
    BLR power law index ($\beta_{BLR}$) & $4$ \\
    \hline
\end{tabular}
\end{table}

\begin{figure*}
\resizebox{\hsize}{!}{\includegraphics[clip=true]{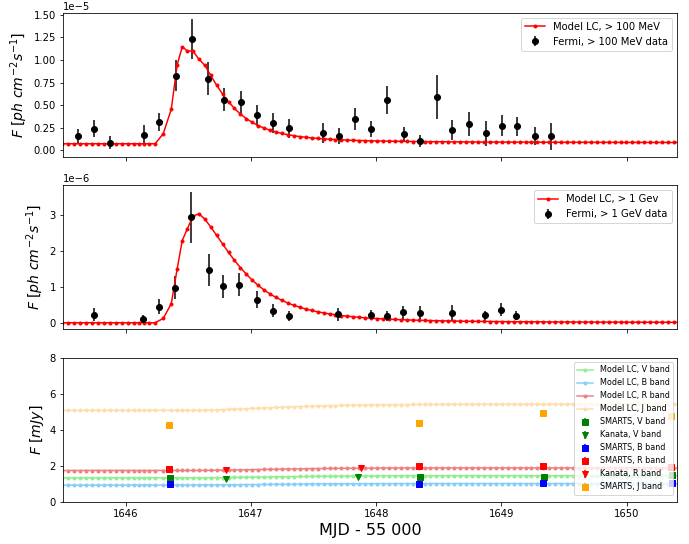}}
\caption{\footnotesize
Simulated light curves in different bands ($E > 1 \ GeV$, $E > 100 \ MeV$, optical and infrared) from the emissions of both blob 1 and blob 2.}
\label{fig:lcmodel}
\end{figure*}

The scenario based on one stationary and one moving blob yields a satisfactory reproduction of the SED and multi-wavelength light curves, as shown in Figure \ref{fig:lcmodel}. The acceleration of blob 2 up to a distance not far above the inner radius of the BLR, leads to a rapid increase of the EIC emission and a steep rise in the high-energy light curves. The following slow decay of the light curve echoes the decrease in the BLR photon density profile at large radii. 
The dominating synchrotron emission from the stationary blob 1 leads to a constant flux at low energies.

The set of parameters proposed for this specific application to the flare data from 3C\,279 points to a large difference in the extension, particle density and magnetic field strength between the two blobs. One might imagine a steady emission region concentrated around the inner spine of the jet close to its base, and a moving blob filling out the full width of the jet, but the viability of such an interpretation requires some further examination. 

In any case, the choice of parameters needs to ensure that the synchrotron emission from the moving blob in the low-energy bands and its SSC emission in the X-ray band remain subdominant compared to emission from blob 1. Alternative parameter sets may be possible and are under investigation.  

The present scenario shows an increase in the X-ray emission from blob 2 at the end of the flare decay phase, due to a reduction in IC cooling. The current set of parameters ensures that this increase is consistent with constraints from existing X-ray data after the flare \citep{Paliya2016}, although a more thorough modelling of the full data set after the flare has not been attempted.
The large difference between the two parameter sets could be reduced if one admitted that emission from blob 2 became negligible after the flare decay, e.g.\ due to a drop in continuous electron injection. 

More generally, this first application of our scenario clearly illustrates the fact that the displacement of a blob in the jet can have an important observable effect on its emission on short observer time scales (see \citet{Finke_2016}).
Notwithstanding a necessary further investigation of the model parameters, the scenario provides a natural explanation for the occurrence of orphan flares and of highly asymmetric flare profiles in FSRQ type blazars.

\bibliographystyle{aa}
\bibliography{bibliography}

\end{document}